# Structure and Dynamics of Superconducting $Na_xCoO_2$ Hydrate and Its Unhydrated Analog

J. W. Lynn[1], Q. Huang[1], C. M. Brown[1,2], V. L. Miller[3], M.L. Foo[3], R.E. Schaak[3], C. Y. Jones,[1] E. A. Mackey[4], and R. J. Cava[3,5]

[1]NIST Center for Neutron Research, National Institute of Standards and Technology, Gaithersburg, MD 20899-8562

[2]Department of Materials Engineering, University of Maryland, College Park, MD 20742

[3]Department of Chemistry, Princeton University, Princeton, NJ 08544

[4]Analytical Chemistry Division, National Institute of Standards and Technology, Gaithersburg, MD 20899-8562

[5]Princeton Materials Institute, Princeton University, Princeton, NJ 08544

Abstract

Neutron scattering has been used to investigate the crystal structure and lattice dynamics of superconducting $Na_{0.3}CoO_2 \cdot 1.4(H/D)_2O$, and the "parent" $Na_{0.3}CoO_2$ material. The structure of $Na_{0.3}CoO_2$ consists of alternate layers of $CoO_2$ and Na and is the same as the structure at higher Na concentrations. For the superconductor, the water forms two additional layers between the Na and $CoO_2$, increasing the *c*-axis lattice parameter of the hexagonal $P6_3/mmc$ space group from 11.16 Å to 19.5 Å. The Na ions are found to occupy a different configuration from the parent compound, while the water forms a structure that replicates the structure of ice. Both types of sites are only partially occupied. The $CoO_2$ layer in these structures is robust, on the other hand, and we find a strong inverse correlation between the $CoO_2$ layer thickness and the superconducting transition temperature ($T_C$ increases with decreasing thickness). The phonon density-of-states for $Na_{0.3}CoO_2$ exhibits distinct acoustic and optic bands, with a high-energy cutoff of ~100 meV. The lattice dynamical scattering for the superconductor is dominated by the hydrogen modes, with librational and bending modes that are quite similar to ice, supporting the structural model that the water intercalates and forms ice-like layers in the superconductor.

## I. Introduction

Cobalt oxide systems have been attracting increased attention recently because of their interesting magnetic and thermoelectric properties, as well as for possible analogies to colossal magnetoresistive manganite materials, or high superconducting transition temperature cuprate oxides. For the $Na_xCoO_2$ system of particular interest here, the spin entropy has been found to play an essential role[1] in the dramatically enhanced thermopower[2] for large sodium content (x ~ 0.7), while evidence of magnetic ordering has been observed in this family of materials.[3] On the other hand, the recent discovery[4] of superconductivity in hydrated $Na_xCoO_2$–the first superconductor where the presence of water is critical to superconductivity—has been of particular interest with regard to the superconducting cuprates. This is a layered system where the $Co^{4+}$ ions are in the low-spin state and carry S=1/2, just like the cuprates. In addition, the dependence of the superconducting transition on band filling is reminiscent[5] of the cuprates So far the highest superconducting transition is 5 K, but they are strongly anisotropic type-II materials ($\kappa \sim 10^2$),[6-10] again, similar to the cuprates. These observations suggest that this may be the first new class of "high-$T_C$" superconductors since the discovery of the cuprates over seventeen years ago. Of course, the nature and mechanism of superconducting pairing in this new class of materials is in the early stages of being addressed. The appropriate underlying model may be a Mott insulator in two-dimensions,[11-15] with S=1/2 where quantum fluctuations are optimal.[16] The Co spins would then play a critical role in forming Cooper pairs that might have triplet symmetry[17] as in $Sr_2RuO_4$ or $d$-wave symmetry as in the cuprates. On the other hand, the traditional electron-phonon interaction may be establishing conventional s-wave pairing,[18] with the possibility that the anharmonic motion of the hydrogen and oxygen ions might be playing a role in enhancing the superconducting properties, in a manner similar to $MgB_2$.[19]

In the present article we investigate both the crystal structure and lattice dynamics for three different samples of the superconducting hydrate, and compare the behavior with the related non-superconducting compound $Na_{0.3}CoO_2$. The complex chemistry of the superconducting phase requires unconventional sample handling and data analysis to obtain its important structural and dynamical properties.

## II. Experimental Details

The $Na_{0.3}CoO_2 \cdot 1.4H_2O$ (and $D_2O$ based) polycrystalline samples were prepared by chemically deintercalating sodium from $Na_{0.7}CoO_2$ using bromine as an oxidizing agent. The Na and water contents were chosen to optimize the superconducting transition temperature, and further details of the sample preparation are given elsewhere.[5,20] The $Na_{0.7}CoO_2$ was prepared from $Na_2CO_3$ and $Co_3O_4$ heated overnight in $O_2$ at 800 ºC. A 10% molar excess of $Na_2CO_3$ was added to compensate for loss due to volatilization. For the sample with the $T_C$ of 4.5K, one gram of $Na_{0.7}CoO_2$ was stirred in 40 mL of a $Br_2$ solution in acetonitrile at room temperature for five days. A bromine concentration representing a molar excess of 40× relative to the amount that would theoretically be needed to remove all of the sodium from $Na_{0.7}CoO_2$ was employed. For the sample with the $T_C$ of 2.5K, 1 g of $Na_{0.7}CoO_2$ was treated in 40 mL of 6M $Br_2$ in acetonitrile at ambient temperature for 1 day to deintercalate sodium (2). The products were washed several times with acetonitrile and then water (or $D_2O$), and then dried briefly under ambient conditions.

Powder x-ray diffraction with Cu Kα radiation was employed to initially characterize the samples. The superconducting transition temperatures were measured with a SQUID magnetometer. Four samples were prepared for the neutron measurements, all with a nominal sodium content of x=0.3. One sample contained no water, one was fully deuterated, one contained a mixture of D and H, and one contained no D to maximize the (incoherent) inelastic scattering. The neutron samples were approximately 1 gram in mass, and were loaded in cylindrical vanadium sample holders.

The neutron powder diffraction data for $Na_{0.3}CoO_2$ $1.4(D_2O/H_2O)$ were collected using the BT-1 high-resolution powder diffractometer at the NIST Center for Neutron Research, employing Cu (311) and Ge(311) monochromators to produce monochromatic neutron beams of wavelength 1.5401 Å and 2.0775 Å, respectively. Collimators with horizontal divergences of 15´, 20´, and 7´ full-width-at-half-maximum (FWHM) were used before and after the monochromator, and after the sample, respectively. The intensities were measured in steps of 0.05° in the 2θ range 3°-168°. Data were collected at a variety of temperatures from 300 K to 1.5 K to determine the crystal structure and explore the possibility of magnetic ordering or structural transitions. A top-loading helium cryostat was employed for the cryogenics. The structural parameters were refined using the program GSAS,[21] and the neutron scattering amplitudes used in the refinements were 0.363, 0.253, 0.667, -0.374, and 0.581 ($\times 10^{-12}$ cm) for Na, Co, D, H, and O, respectively. Additional data were obtained on the high-intensity BT-7 triple axis spectrometer to explore possible magnetic scattering in more detail, and to monitor the freeze-drying of the hydrated samples to remove the free water. For these measurements, a pyrolytic graphite PG(002) double monochromator was used at a wavelength of 2.46 Å with a PG filter to suppress higher order neutron wavelength contaminations, 52´ FWHM collimation after the sample, and a PG(002) analyzer set at the elastic position (providing an energy resolution of 1 meV). Some data were also collected on the BT-9 triple-axis spectrometer, with a wavelength of 2.35 Å and similar collimations, analyzer, and filter.

Inelastic neutron scattering measurements were undertaken to determine the generalized phonon density-of-states (GPDOS), which is the phonon density-of-states weighted by the neutron cross section of each ion divided by its mass. The data were collected from 1.5 K to 270 K on the filter analyzer neutron spectrometer (FANS) for energies from 28-260 meV, and on the Fermi Chopper Spectrometer for energies from 0.5-40 meV. Additional details of the inelastic experiments and analysis can be found elsewhere.[22]

We also carried out prompt-gamma analysis of the protonated sample used in the inelastic measurements, to determine if this technique could be helpful in establishing the water content in the superconductor independent of the diffraction analysis. The measurements were carried out on the thermal neutron prompt-gamma analysis instrument to avoid any re-thermalization of the incident neutrons by the protons in the sample. To avoid the possibility of losing water from the superconductor, either before or after the freeze-drying process, the sample was constantly keep below room temperature. To accommodate the closed cycle refrigerator used for this propose, which was much larger than a typical (room temperature) sample used for analysis, modifications of the instrument were necessary. The relative ratios of Na to Co and hydrogen to Co were determined to be 0.35(4) Na/Co, and 1.9(2) H/Co. The Na content is within one standard deviation of the nominal amount. However, the hydrogen content is substantially lower than the value determined chemically, or by our measurement of the *c*-axis lattice constant, which is very sensitive to the water content.  Hence in the present case the crystal structure determination is the reliable way to determine the water content in the superconducting phase.  As we show below, the diffraction results indicate a significant amount of impurity phase in the sample, and the prompt-gamma measurement determines the H/Co ratio for the entire sample, averaged over all phases.  The prompt-gamma analysis establishes that the impurity phase contains Co, but does not contain a significant amount of hydrogen.

### III. Experimental Results and Discussion
#### a) Crystal Structure

Figure 1 shows the diffraction pattern for the $Na_{0.30}CoO_2$ parent material.  The points are the observations and the solid curve represents the calculated profile after fitting to the data.  The difference between calculated and observed is shown at the bottom of the figure.  We started with the crystal structure determined at higher sodium content,[23] and we see that a good fit to the data is achieved with this model.  There are no significant impurity phases in the sample.  The crystallographic parameters resulting from the fit are given in Table 1.

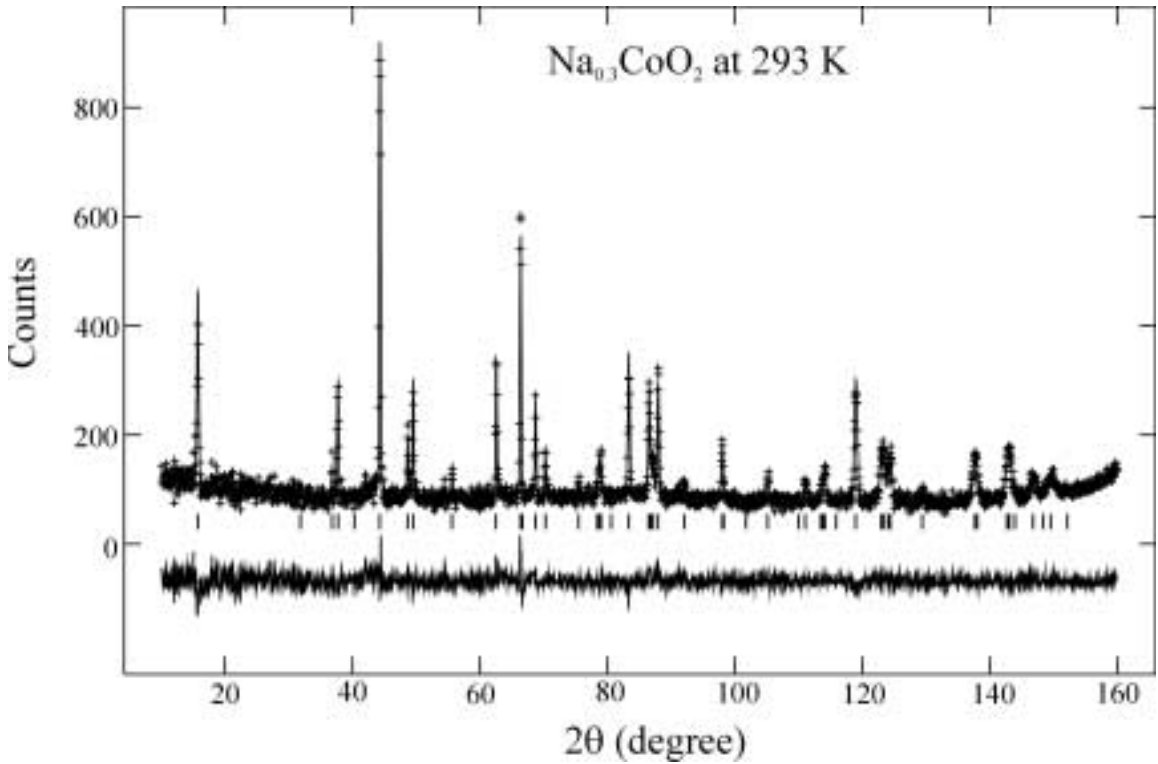

Fig. 1.  (Color online) Diffraction pattern for the $Na_{0.30}CoO_2$ parent material taken with a neutron wavelength of 1.5402 Å, along with the fitted profile (solid curve) and the difference between calculated and observed (bottom).

**Table I.** Structure parameters of $Na_xCoO_2$ 1.4($D_2$/$H_2O$). Space group $P6_3/mmc$ (194), atomic positions: Co: $2a$ (0 0 0 ); O: $4f$ (1/3, 2/3, $z$); Na(1): $2b$(0, 0, 1/4); Na(2): $2c$ (1/3, 2/3, 1/4) for superconducting samples (I and II) , but (2/3, 1/3, 1/4) for sample III; W1: $12k(x, 2x, z)$; W2: $12k(x, 2x, z)$.

| Atom | Parameter | Sample I $Na_xCoO_2$ 1,4($D_2O$) (1.5 K[1)]/237 K[2)]) | Sample II $Na_xCoO_2$ 1.4($D_{1.76}H_{0.24}O$) (235 K[3)]) | Sample III $Na_xCoO_2$ (236 K/293 K) |
|---|---|---|---|---|
| | $T_C$ (K) | 2.5 | 4.5 | 0 |
| | $a$ (Å) | 2.8184(2)/2.8232(1) | 2.8223(3) | 2.8142(1)/2.81233(7) |
| | $c$ (Å) | 19.487(2)/19.530(1) | 19.608(2) | 11.1686(8)/11.2061(4) |
| | $V$ (Å$^3$) | 134.05(2)/134.81(1) | 135.27(3) | 76.60(1)76.757(5) |
| Co | $B$(Å$^2$) | 0.24/1.1(2) | 0.5 | 0.16(7)/0.31(5) |
| O | $z$ | 0.0475(1)/0.0480(1) | 0.0470(2) | 0.0867(2)/0.0865(1) |
| | $B$(Å$^2$) | 0.24(7)/1.6 | 0.69(6) | 0.51(3)/0.64(2) |
| Na(1)[4)] | $n$ | 0.165(11)/0.229(9) | 0.26(3) | 0.17(1)/0.17(1) |
| | $B$(Å$^2$) | 1.0/2.4 | 3.5(9) | 0.5/0.6 |
| Na(2) | $n$ | 0.135(11)/0.071(9) | 0.04(3) | 0.13(1)/0.13(1) |
| | $B$(Å$^2$) | 1.0/2.4 | 3.5(9) | 0.5/0.6 |
| W(1) | $x$ | 0.216(2)/0.211(2) | 0.258(9) | |
| | $y$ | 0.432(5)/0.422(4) | 0.517(18) | |
| | $z$ | 0.1348(6)/0.1357(4) | 0.136(1) | |
| | $n$ | 0.101(2)/0.105(2) | 0.083(9) | |
| | $B$(Å$^2$) | 7.6(2)/8.9(2) | 11.3(6) | |
| W(2) | $x$ | 0.165(2)/0.1595(10) | 0.169(4) | |
| | $y$ | 0.331(3)/0.319(2) | 0.339(8) | |
| | $z$ | -0.1805(4)/-0.1823(3) | 0.179(1) | |
| | $n$ | 0.152(3)/0.141(2) | 0.151(9) | |
| | $B$(Å$^2$) | 7.6(2)/8.9(2) | 11.3(6) | |
| Water content | $y$ | 1.52(4)/1.48(3) | 1.4(2) | |
| | $R_p$(%) | 5.04/5.84 | 3.10 | 5.41/6.21 |
| | $R_{wp}$(%) | 6.65/8.01 | 3.75 | 7.03/7.47 |
| | $\chi^2$ | 3.508/4.549 | 1.448 | 2.232/1.062 |
| | Selected interatomic distances (Å) and angles (degrees). | | | |
| Co-O | ×6 | 1.872(1)/1.880(1) | 1.872(2) | 1.892(1)/1.8911(6) |
| O-Co-O (*A1*) | | 97.68(9)/97.33(8) | 97.85(15) | 96.12(9)/96.07 (4) |
| O-Co-O (*A2*) | | 82.32(9)/82.67(8) | 82.15(15) | 83.88(9)/83.93(4) |
| Na(1)-W(1) | | 2.48(1)/2.460(9) | 2.57(4) | |
| Na(1)-W(2) | | 2.547(2)/2.552(5) | 2.56(2) | |
| Na(2)-W(2) | | 2.32(1)/2.312(8) | | |

1) Data collected before pumping; sample contains 23wt% free $D_2O$ ice.
2) Free $D_2O$ ice was removed by pumping.
3) Sample contains 30wt% free $(D_{0.88}H_{0.12})_2O$ ice, and the relative fraction of H/D was determined from the refinement of the ice.
4) Na content x was constrained to be equal to 0.30.

Figure 2 shows the diffraction pattern collected for the fully deuterated superconducting sample of $Na_{0.3}CoO_2 \cdot 1.4D_2O$, collected at 1.5 K in the superconducting phase. The data show the Bragg peaks from the sample as well as a number of additional strong Bragg peaks that originate from free ice. These ice peaks exhibited some preferred orientation, and the sample was rotated continuously to randomize this scattering so good fits could be achieved. No preferred orientation was observed for the peaks from the superconductor. The data could be refined with the primary and ice phases,[24] and the refinement is also shown in Fig. 2. A good fit to the diffraction pattern was achieved, with the exception of two peaks that are observed at 44° and 48°, along with a few much weaker peaks. The peak at 44° has a *d*-spacing that corresponds to the (forbidden) 007 Bragg peak of the superconductor, and hence this might signal a superlattice structure and associated crystallographic distortion of the superconducting phase. However, none of the other peaks match, and there are no additional forbidden *c*-axis peaks, such as 001, 003, 005, etc., observed. We therefore conclude that these extra peaks originate instead from an impurity phase, of unknown composition. They exhibited little temperature dependence, and there was no significant change in these two peaks from the removal of the free-ice (described below), and they were not modeled in the refinements.

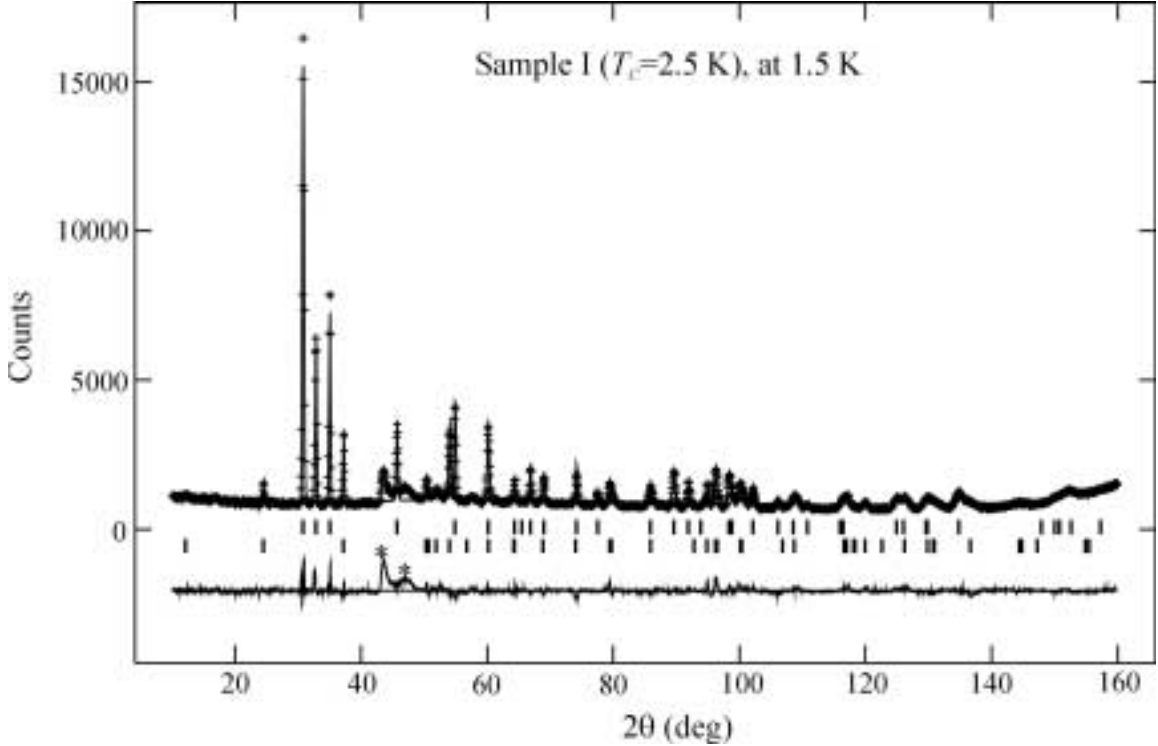

Fig. 2. (Color online) Observed (crosses) and calculated (solid line) intensities for $Na_xCoO_2$ $1.4D_2O$ (sample I) at 1.5 K, taken with a neutron wavelength of 2.0775 Å. The residuals between the observed and calculated intensities are shown in the bottom of figure. The vertical lines indicate possible Bragg peak positions for the superconducting phase (lower) and for free ice (upper). The two impurity peaks are denoted by *.

It is preferable in analyzing the diffraction results to remove the free water if possible, since there is some overlap of peaks. More importantly, to properly identify the inelastic scattering from the superconductor, it is highly desirable to remove the strong scattering from the free water. To remove this free water, we freeze-dried the deuterated sample, using the BT-7 spectrometer to monitor one of the strong ice diffraction peaks. The sample well of the cryostat was continuously pumped while slowly warming the sample from ~120 K to 267 K, just below the freezing point of heavy water, until the intensity of the Bragg peak began to decrease. Pumping was continued until all the free water was removed, which took about nine hours. No significant changes in the primary superconducting phase were observed before and after pumping to remove the free water, but the removal of the ice eliminates any possible interference in the analysis between the ice (with preferred orientation) and the primary phase and the statistical quality of the data for the analysis of the primary phase was improved substantially. The impurity peaks at 44° and 48° did not change significantly when the free water was removed, but the very weak impurity peaks at higher angles increased in size after pumping and became readily visible. This suggests some decomposition (possibly at the surface of the particles) or transformation occurs with the ice removal, but we did not detect any change in the lattice parameters of the superconducting phase, or in its basic composition.

Diffraction data were collected on three samples, a fully deuterated sample with a superconducting transition $T_C$ of 2.5 K, a partially deuterated sample with a $T_C$ of 4.5 K (where the free ice was not removed by pumping), and the "parent" compound that has the same composition but does not contain water and is not superconducting. Data were also collected for the protonated sample, where the excess water was removed by pumping. However, the large incoherent scattering from the hydrogen, optimal for measuring the inelastic scattering, precluded a detailed refinement with the same level of precision as for the deuterated samples. The lattice parameters, in particular the greatly expanded $c$-axis, were essentially identical for all three samples, and did not change significantly before and after the pumping procedure to remove the free ice. However, we have not pursued detailed crystallographic refinements for the protonated sample.

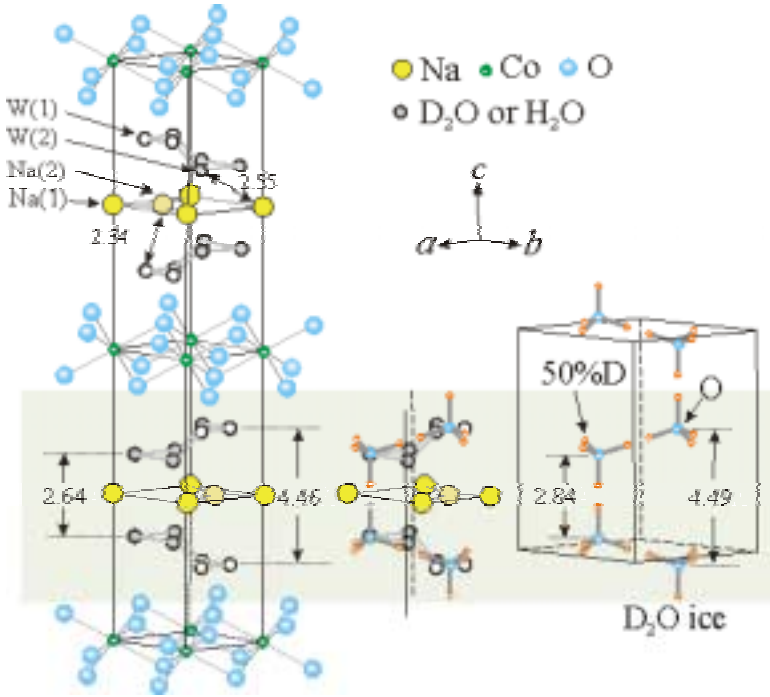

Fig. 3. (Color online) Structure model for the $Na_xCoO_2$ 1.4($D_2$/$H_2O$) superconductor (left). The figure on the right shows the $D_2O$ ice structure.[24] For comparison, the figure in the middle shows the ice molecule superposed on the water block in the structure of the superconducting phase, demonstrating that the dimensions of the water block are close to those found in free ice.

To refine the structures we started with the model proposed by Takada, *et al.*[4] based on x-ray data. The neutron data are of course much more sensitive to the hydrogen, but we found that the sodium positions were also substantially different, and the initial model could not give a good fit to the neutron data. The structure we determined from the neutron measurements is shown in Fig. 3, and consists of alternating layers of $CoO_2$ and Na, but with different Na positions relative to the $CoO_2$ than found in the parent compound and assumed in the x-ray refinements. When the water enters, it intercalates between the $CoO_2$ and Na layers, and the basic elements of the structure are the same; a robust $CoO_2$ layer separated by a trilayer of water/Na/water. The dominant effect is to expand the *c*-axis lattice parameter, from just over 11 Å without water, to almost 20 Å when the system becomes superconducting. Our structure differs from the previous structure determined by x-rays[4] in that we find that the Na ions are displaced compared to the parent compound, while the water in the system exhibits the basic structure that ice possesses. The displacement of the Na is required in order to accommodate these water positions. The final crystallographic results of the refinements are given in Table 1 for all three samples.

Since the model for the structure differs substantially from the model based on x-ray data, we make a detailed comparison of the structures in Fig. 4. The parent material has the Na(2) position at (2/3,1/3,1/4), but we find that the intercalation of the water shifts this to (1/3, 2/3,1/4) relative to the $CoO_2$ layer. Interestingly, the x-ray scattering patterns for these two structures are practically identical, while the neutron patterns are quite different for the two models, and the observed neutron diffraction pattern makes it is clear that the model based on the x-ray data does not explain the data.

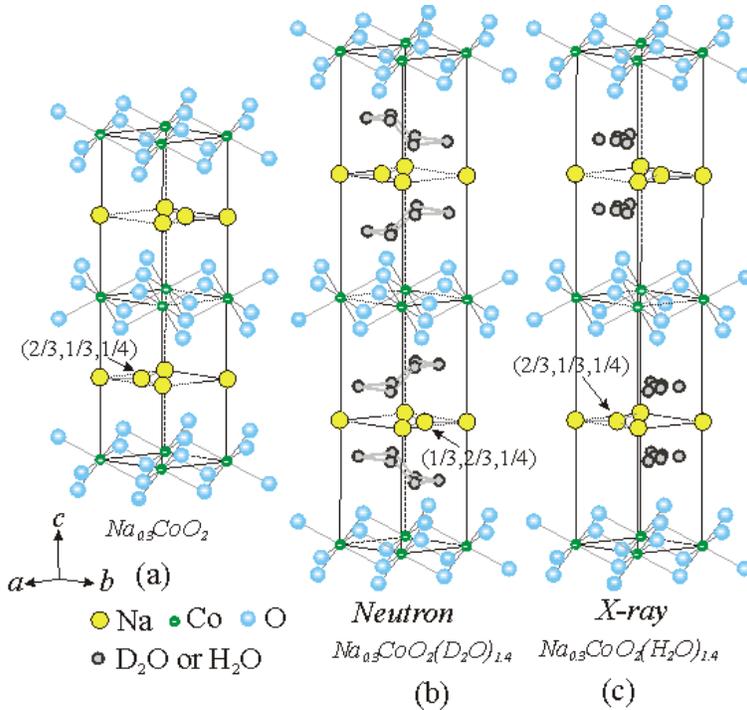

Fig 4. (Color online) Structures of (a) $Na_{0.3}CoO_2$, (b) the superconducting phase model used in the present work and (c) the model reported in ref. [4]. Note that the positions for the Na and water centers are different in (b) and (c).

The neutron data also determine the water positions, and the overall structure is given in Fig. 3. One of the technical problems in carrying out the refinements is that the Na and water positions are both only fractionally occupied. The sodium content is only 0.3, and this is shared between two sites. In the refinements these separate occupancies become correlated with the thermal parameters, hampering the information that can be extracted separately. The Na content from the refinements was found to be 0.28 for sample I, and 0.30 for samples II and III, by fixing the thermal parameters to reasonable values and not allowing them to vary. Since these values are within error of the nominal values determined by the chemistry, in the final refinements we constrained the total Na content to be the content determined chemically (=0.30), and then refined the occupancies and thermal factors. Note that the Na(2) site has the lower occupancy in all cases, but goes from being essentially unoccupied (0.04(3)) for the highest $T_C$ sample, to a significant value for the fully deuterated, lower $T_C$ sample. The data also suggest that the relative occupancy of the two Na sites may change when the free water was removed. However, given the correlation with the thermal factors, this suggestion is tentative, and exploring the influence of the total Na content and the relative site occupancies is an area where further work is indicated.

The water centers, on the other hand, are well determined in the neutron refinements by the oxygen. The positions of the hydrogen are randomized as they are in ice (leading to the famous excess entropy problem of ice) and specific hydrogen occupancies could not be determined. Again, the small fractional occupancies of these sites makes these determinations challenging, but we have found that the crystallographic structure of the $(H/D)_2O$ parallels the structure of ice, as shown in Fig. 3. The ice structure is shown in the right, while the inset in the middle shows the ice model with the Na layer. Both sets of sites are fractionally occupied. Our refinements suggest that the oxygen in the water has a tendency to shift toward the Na when that site is occupied, with the hydrogen then tending to bond to the oxygen in cobalt layer, rendering a subtle distortion of the $D_2O$ layers from the ideal ice structure.

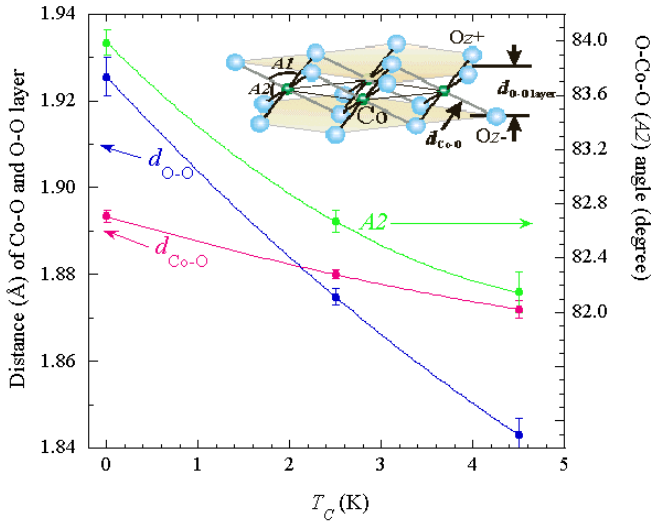

Fig. 5. (Color online) Interatomic distances for Co-O in the $CoO_2$ layer, and the O-Co-O bond angle, as a function of the superconducting transition temperature $T_C$. Note the strong dependence of $T_C$ on the bond angle A2 (defined in the inset), which determines the $CoO_2$ layer thickness.

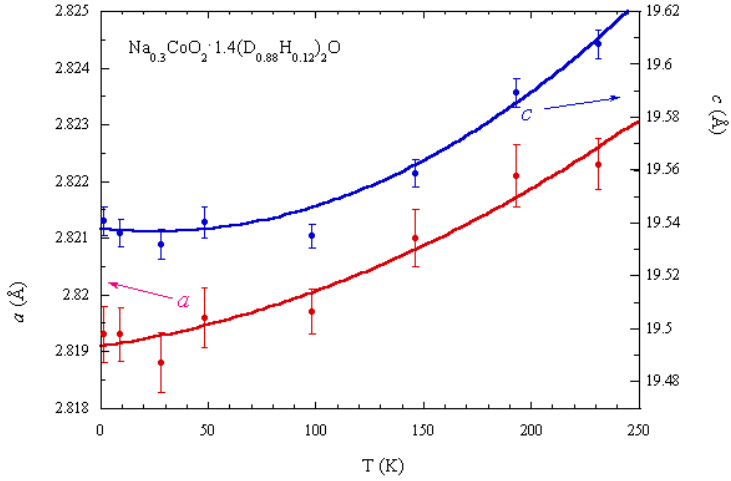

Fig. 6. (Color online) Lattice parameters versus temperature for the superconductor. The lattice expands in both directions, in contrast to the parent compound where only the *c*-axis exhibits a significant expansion (at higher Na content).[25]

The $CoO_2$ layer, on the other hand, is robust, and the diffraction results reveal an interesting correlation between the structure of this layer and the superconducting transition temperature. The Co-O distance is shown in Fig. 5, and there is a modest decrease with increasing $T_C$, which from a crystal chemistry point of view would suggest some electron transfer off the cobalt. The bond angle for O-Co-O, on the other hand, varies quite substantially with $T_C$. This means that the layer thickness (or alternatively, the bond angle) exhibits a large change, from 1.93 Å for $T_C = 0$ K to 1.84 Å for $T_C = 4.5$ K.

Figure 6 shows the measured lattice parameters as a function of temperature. With increasing temperature there is an expansion in both the *a* and *c* lattice parameters, in contrast to the parent material at higher Na content where only the c-axis shows a significant expansion with increasing T.[25] In the superconducting phase, we note that there is a suggestion that there may be a small increase in *a* when entering the superconducting state. However, this is well within the combined errors, and much higher accuracy data would be required to determine if there is any real effect with temperature. We remark that no lattice parameter anomaly has been observed in diffraction for the highest $T_C$ (anharmonic) electron-phonon superconductor, $MgB_2$, and has only very recently been observed in high-resolution dilatometry.[26]

Finally, we searched on BT-7 and BT-9 for evidence of long range magnetic order in the parent compound, in the range of scattering angles from 3° to 65°. Without Na, each Co should be +4, which is in the low spin state ($S = \frac{1}{2}$). These Co spins occupy a triangular lattice, which would be frustrated if the lattice were fully occupied and the in-plane nearest-neighbor exchange interactions were antiferromagnetic. The addition of a Na ion changes the (formal) Co valence to +3, which is non-magnetic. Quantum effects are expected to reduce the $S = \frac{1}{2}$ moment by a factor ~2/3, so on the average we might expect 4/9 $\mu_B$ on the Co sites if the spins are fully ordered. The magnetic Bragg peaks in diffraction are proportional to the square of the average ordered moment, and such a small ordered moment would be difficult to observe in powder diffraction—certainly more difficult than for the Cu moments in the undoped cuprates. Data were collected at 1.5 K, 7 K, 10 K, 20 K, and 40 K for the $Na_{0.3}CoO_2$ sample. Subtracting the high temperature from the lower temperature data did not reveal an indication of magnetic ordering at any of the commensurate (structural) Bragg peaks. Away form the structural Bragg peaks, the only additional peak observed occurred at a position that would correspond to (2/3,0,0), which is a reasonable position to expect a magnetic peak in this triangular spin system. However, this peak was measured from 1.5 K to 250 K, and did not exhibit any temperature dependence, so it does not seem to be related to any kind of magnetic order. It could be an indication of charge order on the cobalt lattice or Na ordering in the system, and this is a possibility that should be explored when single crystals become available.

### b) Dynamics

The generalized phonon density-of-states was measured using the filter analyzer spectrometer and the Fermi chopper spectrometer, over the energy range from 0.5 meV to 260 meV. The data for the three samples investigated, superconducting $Na_xCoO_2 \cdot 1.4 (H_2O)$, $Na_xCoO_2 \cdot 1.4 \cdot (D_2O)$, and the "parent" $Na_{0.3}CoO_2$, are shown in Fig. 7. Both hydrated/deuterate samples used for the inelastic measurements were those without free ice, prepared by freeze drying. For $Na_{0.3}CoO_2$, in the low energy range we have the acoustic phonons that band in the energy range from 0-40 meV, and then the optic modes that band in the range from 50-100 meV. Similar features are observed for the hydrated samples, indicating that the phonons associated with the $CoO_2$ layer and the Na layer are similar in the two materials, as might be expected. For the $Na_{0.3}CoO_2$ material the one-phonon density of states cuts off at ~100 meV, with the additional scattering observed at higher energies originating from multiphonon scattering.

For the $H_2O$ superconducting sample, the scattering is much stronger in the energy range from 50–120 meV than for the parent compound, indicating that this scattering is dominated by the very strong hydrogen scattering. The features in this scattering are in fact quite similar to the scattering observed in free $H_2O$,[27] which originates in this energy range from the librational modes of the water. This is confirmed by the results for the deuterated sample, whose scattering in this energy range closely follows the scattering for the protonated sample after scaling the energies by $(m_H/m_D)^{1/2}$ to account for the change in isotopic mass.

Two data sets were taken for the protonated sample, at 4.2 K and 125 K. At lower energies the data are essentially identical, indicating that there are no substantial changes between these temperatures. In the high energy regime (above ~125 meV), on the other hand, we find several peaks at low temperatures, and these correspond relatively well to the observed bending vibrations of $H_2O$. At elevated temperatures the scattering from these vibrations should be damped by the Debye-Waller factor, and we indeed found that the peaks in this scattering are not evident at 125 K, and all that remains appears to be a multiphonon contribution. We remark that for the filter analyzer type of measurement, the energy and wave vector regimes are coupled together in the measurement, so that higher energies measure at higher wave vectors. With the large and strongly temperature dependent mean-square displacement of hydrogen, at elevated temperatures single-phonon scattering then is strongly suppressed at higher energies/wave vectors.

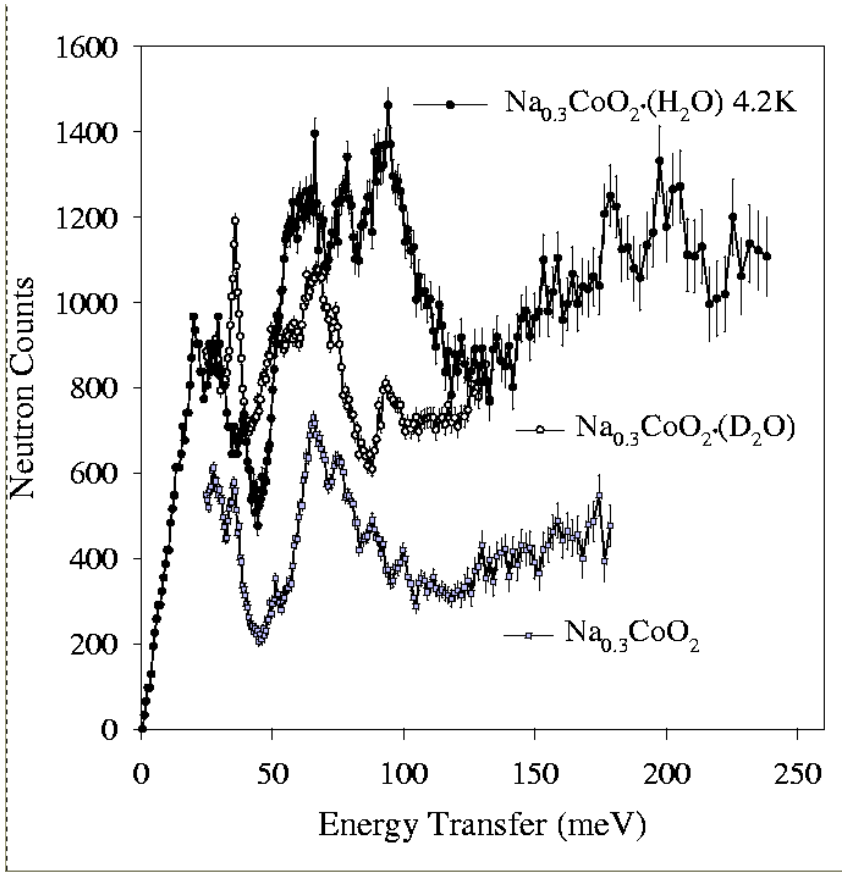

Fig. 7. Measurements of the generalized phonon density-of-states for the parent compound, the deuterated superconductor, and the protonated superconductor. For $Na_{0.3}CoO_2$ the excitations divide into an acoustic band and an optic band, with a one-phonon density-of-states cutoff around 100 meV; at higher energies the scattering is multiphonon in origin. For the protonated superconducting sample there is additional strong scattering in the 50-120 meV range, which matches relatively well to the librational modes of ice,[27] while at higher energies there are bending modes. The intensities of these bending modes have a significant temperature dependence due to Debye-Waller effects, while below ~150 meV temperature has little effect. For the deuterated sample the positions of the hydrogen modes scale with the mass. The sharp peak at 37 meV also agrees with the peak position from ice, indicating that the dynamics of the water in the superconductor is quite similar to the dynamics (and structure) of ice.

Finally, we note the rather strong peak in the scattering at 37 meV, which is particularly evident in the $Na_{0.3}CoO_2 \cdot 1.4D_2O$ sample. This sharp peak is in excellent agreement with the peak observed in $D_2O$ ice [27]. The rather remarkable similarities between the lattice dynamical density-of-states of the superconducting compounds measured here and ice strongly support our structural model, where the intercalated ice in the superconductor forms a structure that mimics the ice structure to a good approximation.

In summary, the neutron diffraction results have revealed that the Na positions in the superconductor compound differ from the parent compound, while the water forms layers with a structure that closely resembles ice. This is the first material where water plays an essential role in the superconductivity, and the nature of the superconducting pairing mechanism is a question of central interest. The $Co^{+4}$ ions have an unpaired spin, and these spins could play a central role in forming the pairs, with a *d*-wave symmetry as in the cuprates. With a relatively low $T_C$ of 5 K, a conventional electron-phonon interaction could also easily produce the pairing, with the interesting possibility that the anharmonic motion of the hydrogen and oxygen ions could play a role in a manner similar to $MgB_2$. These questions can only be answered by further studies, but an essential step is to determine and understand the crystal structure and lattice dynamics that have been elucidated in the present study.

**Acknowledgments**

We thank Terry Udovic for his generous assistance with some of the inelastic neutron measurements, and Dan Neumann for valuable discussions. The work at Princeton University was supported by NSF grants DMR-0244254 and DMR-0213706, and by DOE-BES grant DE-FG02-98-ER45706. Work at the NCNR was supported in part by the Binational Science Foundation, grant 2000073.